\documentclass[conference]{IEEEtran}
\IEEEoverridecommandlockouts

\usepackage{stfloats}
\usepackage{amsmath}
\usepackage{amssymb,latexsym}
\usepackage{graphicx}
\usepackage{color,cite,times}
\usepackage{epstopdf}
\usepackage{algorithmic}
\usepackage[ruled,linesnumbered] {algorithm2e}
\usepackage{booktabs}
\usepackage{bbm} 
\usepackage{url}
\usepackage{fancyhdr}
\usepackage{cleveref}
\usepackage{verbatim}
\usepackage{multirow}
\usepackage{makecell}
\usepackage{graphicx}
\usepackage{indentfirst}
\usepackage{cases}
\newcommand{\RNum}[1]{\uppercase\expandafter{\romannumeral #1\relax}}
\usepackage{colortbl} 
\usepackage{extarrows}
\usepackage{threeparttable}
\usepackage{diagbox}
\usepackage{pifont}
\usepackage{cancel}
\usepackage{framed}
\usepackage{overpic}
\usepackage{mathtools} 

\usepackage{subfigure}

\usepackage[framemethod=tikz]{mdframed}
\usepackage{lipsum}
\usepackage{flushend} 

\newcommand{\bsm}{\boldsymbol}

\newcommand{\mbf}{\mathbf}

\newtheorem{theo}{Theorem}

\definecolor{gray}{RGB}{192,192,192}

\begin{document}
\title{Score-Based Turbo Message Passing for Plug-and-Play Compressive Image Recovery}


\author{\IEEEauthorblockN{Chang Cai$^*$, Xiaojun Yuan$^\dag$, and Ying-Jun Angela Zhang$^*$}
	
	\IEEEauthorblockA{$^*$Department of Information Engineering, The Chinese University of Hong Kong, Shatin, N.T., Hong Kong SAR\\
		$^\dag$National Key Lab. of Wireless Commun., Uni. of Electronic Sci. and Tech. of China, Chengdu, China \\
		Email: cc021@ie.cuhk.edu.hk,
		xjyuan@uestc.edu.cn,
		yjzhang@ie.cuhk.edu.hk
		\thanks{
			This work is supported in part by the General Research Fund (project number 14202421, 14214122, 14202723, 14207624), Area of Excellence Scheme grant (project number AoE/E-601/22-R), and NSFC/RGC Collaborative Research Scheme (project number CRS\_HKUST603/22, CRS\_HKU702/24), all from the Research Grants Council of Hong Kong.}
	}
}
\maketitle

\begin{abstract}
	Message passing algorithms have been tailored for compressive imaging applications by plugging in different types of off-the-shelf image denoisers.
	These off-the-shelf denoisers mostly rely on some generic or hand-crafted priors for denoising.
	Due to their insufficient accuracy in capturing the true image prior, these methods often fail to produce satisfactory results, especially in highly underdetermined scenarios.
	On the other hand, score-based generative modeling offers a promising way to accurately characterize the sophisticated image distribution.
	In this paper, by exploiting the close relation between score-based modeling and empirical Bayes-optimal denoising, we devise a message passing framework that integrates a score-based minimum mean squared error (MMSE) denoiser for compressive image recovery.
	Experiments on the FFHQ dataset demonstrate that our method strikes a significantly better performance-complexity tradeoff than conventional message passing, regularized linear regression, and score-based posterior sampling baselines.
	Remarkably, our method typically converges in fewer than 20 neural function evaluations (NFEs).
\end{abstract}

\section{Introduction}
Consider the problem of recovering an image $\mbf{x} \in \mathbb{R}^{N}$ from noisy linear measurements $\mbf{y} \in \mathbb{R}^{M}$ of the form
\begin{align} \label{linear_inverse_problem}
	\mbf{y} = \mbf{A}\mbf{x} + \mbf{n},
\end{align}
where $\mbf{A} \in \mathbb{R}^{M \times N}$ is a known measurement matrix, and $\mbf{n} \in \mathbb{R}^{M}$ is an additive white Gaussian noise (AWGN) with zero mean and covariance $\delta_0^2 \mbf{I}$.
This problem arises in different imaging domains, such as magnetic resonance imaging, synthetic aperture radar, and computed tomography.
In these applications, it is often the case of $M \ll N$, making the linear inverse problem underdetermined.
Therefore, it is essential to incorporate prior knowledge of $\mbf{x}$ into the estimation process.
This problem is broadly referred to as \textit{compressive image recovery} in the literature \cite{from_denoising_to_cs}.

Message passing algorithms \cite{from_denoising_to_cs, ci_amp2015tsp, xue2017access} stand as a powerful tool for compressive image recovery. 
For example, denoising-based approximate message passing (D-AMP) \cite{from_denoising_to_cs} and denoising-based turbo compressive sensing (D-Turbo-CS) \cite{xue2017access} are able to incorporate a variety of off-the-shelf image denoisers.
Popular choices include the total variation (TV) denoiser, the SURE-LET denoiser \cite{sure-let}, the BM3D denoiser \cite{bm3d}, etc.
These off-the-shelf denoisers mostly rely on non-parametric models for denoising, which do not require the hypothesis of a statistical model for the noiseless image.
Consequently, the above-mentioned plug-and-play methods can result in sub-optimal performance, as they fail to fully exploit the strong prior inherent in image data.
Some recent works \cite{metzler2017learned_damp, versatile} incorporate either a denoising CNN (DnCNN) or an unrolled network in place of the off-the-shelf denoiser.

On the other hand, score-based generative models \cite{score_sde} have achieved unparalleled success in generating high-quality samples across various domains. 
This evidences their remarkable capabilities in capturing the intricacies of high-dimensional data distribution.
During training, score-based generative models perturb the data with random Gaussian noise of different magnitudes.
They learn the data prior by matching the gradient of the log density of the perturbed data, known as \textit{score function}.
The score function is then applied at each and every reverse diffusion step for sample generation.
Score-based generative models can also be leveraged for solving inverse problems without task-specific training \cite{kawar2021snips, dps, diffpir}.
These methods obtain a prior-related term (i.e., the score function) from the score-based generative network and a likelihood term from the degradation model.
The two terms are combined to form the posterior term via Bayes' theorem, enabling solving inverse problems stochastically by sampling from the posterior distribution.
However, these methods typically require traversing the entire reverse diffusion process, necessitating hundreds or even thousands of neural function evaluations (NFEs), which makes them computationally expensive.

In this paper, we devise a score-based message passing algorithm for compressive imaging. 
This approach inherits the fast convergence speed of message passing and, meanwhile, fully leverages the image prior through the integration of score networks.
We exploit the fact that, at each iteration, message passing algorithms construct a pseudo AWGN observation of the ground-truth image.
We repurpose the score function for denoising the AWGN observation through Tweedie's formula \cite{efron2011tweedie}, achieving per-iteration minimum mean squared error (MMSE) performance with a single call of the score network. 
We further relate the posterior variance calculation to the second-order generalization of Tweedie's formula, which depends on the Hessian of the log density (i.e., the second-order score function) \cite{lu2022maximum}.
We propose matching the second-order score function to obtain the posterior variance, facilitating message updates in a turbo-type message passing algorithm,
hence the name \textit{score-based turbo message passing} (STMP).
Experimental results on the FFHQ dataset \cite{ffhq} demonstrate that STMP strikes a much better tradeoff between performance and complexity than conventional message passing, regularized linear regression, and score-based posterior sampling schemes.

\begin{figure}[t]
	\vskip 0.05in
	\begin{center}
		\centerline{\includegraphics[width=1\columnwidth]{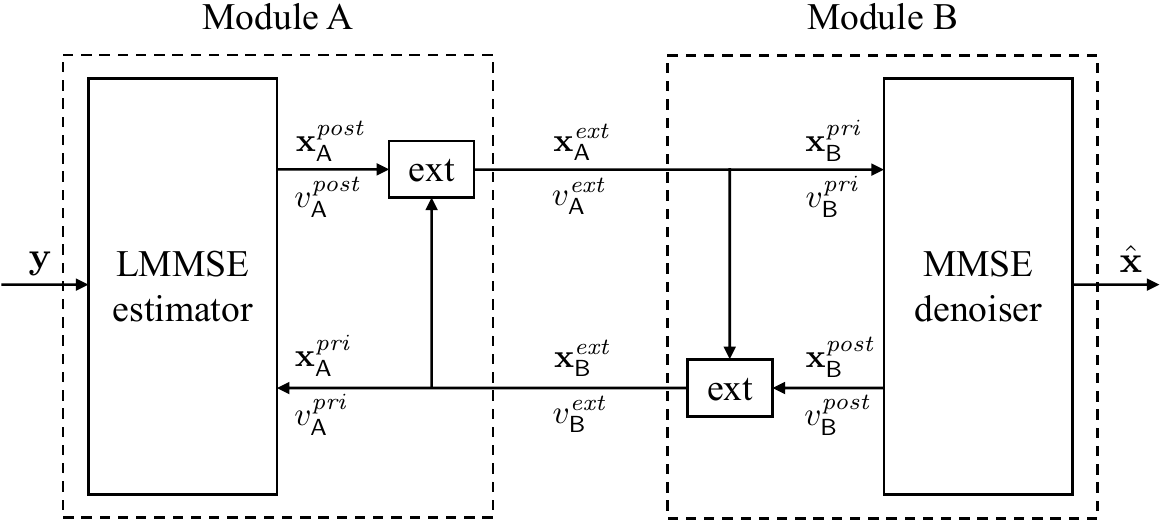}}
		\caption{A modularized representation of the Turbo-CS framework \cite{turbo_cs}.}
		\label{turbo_cs_module_new}
	\end{center}
	\vskip -0.25in
\end{figure}

\section{Review of Turbo-CS Framework}
The Turbo-CS framework \cite{turbo_cs}, also known as vector AMP (VAMP) \cite{vamp}, is a well-established method for approximating the MMSE estimator $\mathbb{E}[\mbf{x}|\mbf{y}]$.
This framework can be derived using various approaches such as expectation consistent (EC) approximation and expectation propagation (EP).
As illustrated in Fig. \ref{turbo_cs_module_new}, Turbo-CS consists of two modules and admits a modularized interpretation that resembles the turbo decoder.
In particular, module A is a linear MMSE (LMMSE) estimator of $\mbf{x}$ exploiting the likelihood $p(\mbf{y}|\mbf{x})$ and the messages from module B.
Module B is an MMSE denoiser incorporating the prior distribution $p(\mbf{x})$ and the messages from module A.
The two modules are executed iteratively to refine the estimate.

Module A calculates the posterior estimate of $\mbf{x}$ by combining the likelihood $p(\mbf{y}|\mbf{x}) = \mathcal{N}(\mbf{y}; \mbf{A}\mbf{x}, \delta_0^2\mbf{I})$ with the pseudo Gaussian prior $\mathcal{N}\big(\mbf{x}; \mbf{x}_{\sf A}^{\textit{pri}}, v_{\sf A}^{\textit{pri}}\mbf{I}\big)$ as 
\begin{align}
	\mathcal{N}\big(\mbf{x}; \mbf{x}_{\sf A}^{\textit{post}}, v_{\sf A}^{\textit{post}}\mbf{I}\big) \propto \mathcal{N}(\mbf{y}; \mbf{A}\mbf{x}, \delta_0^2\mbf{I}) \mathcal{N}\big(\mbf{x}; \mbf{x}_{\sf A}^{\textit{pri}}, v_{\sf A}^{\textit{pri}}\mbf{I}\big),
\end{align}
where
\begin{align}
	\mbf{x}_{\sf A}^{\textit{post}} &= \mbf{x}_{\sf A}^\textit{pri} + v_{\sf A}^\textit{pri}\mbf{A}^{\sf T} \big(v_{\sf A}^\textit{pri}\mbf{A}\mbf{A}^{\sf T} + \delta_0^2\mbf{I}\big)^{-1} \big(\mbf{y} - \mbf{A} \mbf{x}_{\sf A}^\textit{pri}\big), \label{lmmse_mean}\\
	v_{\sf A}^\textit{post} &= v_{\sf A}^\textit{pri}- \frac{(v_{\sf A}^\textit{pri})^2}{N} \mathrm{tr} \left(\mbf{A}^{\sf T} \big(v_{\sf A}^\textit{pri}\mbf{A} \mbf{A}^{\sf T} + \delta_0^2 \mbf{I}\big)^{-1}\mbf{A}\right). \label{lmmse_variance}
\end{align}
Eqns. \eqref{lmmse_mean} and \eqref{lmmse_variance} correspond to the posterior mean and variance from the LMMSE estimator, respectively.
The posterior distribution cannot be directly passed to module B due to the correlation issue \cite{oamp}.
Instead, we compute the extrinsic message by excluding the contribution of the prior message as
\begin{align}
	\mathcal{N}\big(\mbf{x}; \mbf{x}_{\sf A}^{\textit{ext}}, v_{\sf A}^{\textit{ext}}\mbf{I}\big) \propto \frac{\mathcal{N}\big(\mbf{x}; \mbf{x}_{\sf A}^{\textit{post}}, v_{\sf A}^{\textit{post}}\mbf{I}\big)}{\mathcal{N}\big(\mbf{x}; \mbf{x}_{\sf A}^{\textit{pri}}, v_{\sf A}^{\textit{pri}}\mbf{I}\big)},
\end{align}
where the extrinsic variance and mean are respectively given by
\begin{align}
	v_{\sf A}^{\textit{ext}} &= \left(\frac{1}{v_{\sf A}^{\textit{post}}} - \frac{1}{v_{\sf A}^{\textit{pri}}}\right)^{-1}, \\
	\mbf{x}_{\sf A}^{\textit{ext}} &= v_{\sf A}^{\textit{ext}} \left(\frac{\mbf{x}_{\sf A}^{\textit{post}}}{v_{\sf A}^{\textit{post}}} - \frac{\mbf{x}_{\sf A}^{\textit{pri}}}{v_{\sf A}^{\textit{pri}}}\right).
\end{align}

The extrinsic message of module A serves as the input to module B, i.e., $\mathcal{N}\big(\mbf{x}; \mbf{x}_{\sf B}^{\textit{pri}}, v_{\sf B}^{\textit{pri}}\mbf{I}\big) = \mathcal{N}\big(\mbf{x}; \mbf{x}_{\sf A}^{\textit{ext}}, v_{\sf A}^{\textit{ext}}\mbf{I}\big)$.
Module B assumes a pseudo likelihood $p\big(\mbf{x}_{\sf B}^\textit{pri}|\mbf{x}\big) = \mathcal{N}\big(\mbf{x}_{\sf B}^{\textit{pri}}; \mbf{x}, v_{\sf B}^{\textit{pri}}\mbf{I}\big)$.
That is, $\mbf{x}_{\sf B}^\textit{pri}$ is modeled as an AWGN observation of the ground-truth $\mbf{x}$:
\begin{align}
	\mbf{x}_{\sf B}^\textit{pri} = \mbf{x} + \mbf{w}, ~~ \mbf{w} \sim \mathcal{N}\big(\mbf{0}, v_{\sf B}^\textit{pri}\mbf{I}\big). \label{AWGN_observation}
\end{align}
This assumption is justified by the state evolution (SE) analysis in \cite{vamp_se_non_separable}, showing that the error vector $\mbf{x}_{\sf B}^\textit{pri} - \mbf{x}$ is asymptotically i.i.d. Gaussian in the large system limit with right-rotationally invariant $\mbf{A}$.
Module B calculates the posterior estimate of $\mbf{x}$ by combining the prior $p(\mbf{x})$ with the pseudo Gaussian likelihood $p\big(\mbf{x}_{\sf B}^\textit{pri}|\mbf{x}\big) = \mathcal{N}\big(\mbf{x}_{\sf B}^{\textit{pri}}; \mbf{x}, v_{\sf B}^{\textit{pri}}\mbf{I}\big)$ following the EP rule as
\begin{align}
	\mathcal{N}\big(\mbf{x}; \mbf{x}_{\sf B}^{\textit{post}}, v_{\sf B}^{\textit{post}}\mbf{I}\big) = \mathrm{proj}\big[p(\mbf{x}) \mathcal{N}\big(\mbf{x}_{\sf B}^{\textit{pri}}; \mbf{x}, v_{\sf B}^{\textit{pri}}\mbf{I}\big) \big], \label{module_B_EP}
\end{align}
where $\mathrm{proj}[p(\cdot)]$ denotes the projection of $p(\cdot)$ to a Gaussian density with matched first- and second-order moments.
Eqn. \eqref{module_B_EP} can be interpreted as the MMSE denoising under the AWGN observation model \eqref{AWGN_observation}:
\begin{align}
	\mbf{x}_{\sf B}^{\textit{post}} &= \mathbb{E} \big[\mbf{x}\big|\mbf{x}_{\sf B}^{\textit{pri}}\big], \label{post_mean}\\
	v_{\sf B}^\textit{post} &= \frac{1}{N} \mathrm{tr} \big(\mathrm{Cov} \big[\mbf{x}\big|\mbf{x}_{\sf B}^{\textit{pri}}\big] \big). \label{post_variance}
\end{align}
The extrinsic message of module B is calculated similar to that of module A as
\begin{align}
	v_{\sf B}^{\textit{ext}} &= \left(\frac{1}{v_{\sf B}^{\textit{post}}} - \frac{1}{v_{\sf B}^{\textit{pri}}}\right)^{-1}, \\
	\mbf{x}_{\sf B}^{\textit{ext}} &= v_{\sf B}^{\textit{ext}} \left(\frac{\mbf{x}_{\sf B}^{\textit{post}}}{v_{\sf B}^{\textit{post}}} - \frac{\mbf{x}_{\sf B}^{\textit{pri}}}{v_{\sf B}^{\textit{pri}}}\right).
\end{align}
Finally, the extrinsic message of module B flows leftward to module A, serving as the prior message of module A in the next iteration, i.e., $\mathcal{N}\big(\mbf{x}; \mbf{x}_{\sf A}^{\textit{pri}}, v_{\sf A}^{\textit{pri}}\mbf{I}\big) = \mathcal{N}\big(\mbf{x}; \mbf{x}_{\sf B}^{\textit{ext}}, v_{\sf B}^{\textit{ext}}\mbf{I}\big)$.

The main challenge in implementing the Turbo-CS algorithm arises from the MMSE denoising part in module B, primarily due to i) the lack of a tractable prior distribution and ii) the high-dimensional integrals involved in calculating the posterior mean and variance. 
To sidestep these difficulties, various off-the-shelf image denoisers (e.g., TV, SURE-LET \cite{sure-let}, BM3D \cite{bm3d}) are plugged into message passing algorithms in place of the principled MMSE denoiser.
These plug-and-play denoisers typically rely on generic or hand-crafted priors, which could lead to substantial imperfection in compressive imaging applications due to oversimplification and misspecification of the prior distribution.
Motivated by the above, we propose a score-based turbo message passing (STMP) algorithm that exploits the score-based generative models as an accurate and expressive image prior, facilitating compressive image recovery from a Bayesian perspective.

\section{Score-Based MMSE Denoising}
\begin{theo}[Tweedie's formula and its second-order generalization \cite{efron2011tweedie}] \label{tweedie_mean_covariance}
	Consider $\tilde{\mbf{x}}$ to be an AWGN observation of $\mbf{x}$, i.e., $\tilde{\mbf{x}} = \mbf{x} + \mbf{w}$, where the prior distribution $p(\mbf{x})$ and the noise distribution $\mbf{w} \sim \mathcal{N}(\mbf{0}, \sigma^2 \mbf{I})$ are given.
	The posterior mean and covariance are respectively given by
	\begin{align}
		\mathbb{E} \left[\mbf{x}|\tilde{\mbf{x}}\right] &= \tilde{\mbf{x}} + \sigma^2 \nabla_{\tilde{\mbf{x}}} \log p(\tilde{\mbf{x}}), \label{1st_order_tweedie} \\
		\mathrm{Cov} \left[\mbf{x}|\tilde{\mbf{x}}\right] &= \sigma^2 \mbf{I} + \sigma^4 \nabla_{\tilde{\mbf{x}}}^2 \log p(\tilde{\mbf{x}}), \label{2nd_order_tweedie}
	\end{align}
	where $\nabla_{\tilde{\mbf{x}}} \log p(\tilde{\mbf{x}})$ and $\nabla_{\tilde{\mbf{x}}}^2 \log p(\tilde{\mbf{x}})$ denote the first- and second-order score function of the perturbed data, respectively.
\end{theo}

Theorem \ref{tweedie_mean_covariance} establishes the theoretical foundation for using score functions to achieve empirical Bayes-optimal denoising of AWGN observations.
Once the score functions are known, we can apply Theorem \ref{tweedie_mean_covariance} to efficiently implement the MMSE denoiser in \eqref{post_mean} and \eqref{post_variance}.
We discuss how to obtain the first- and second-order score functions via the denoising score matching technique \cite{vincent2011connection, lu2022maximum} in the following.

\section{First- and Second-Order Score Matching}
\subsection{First-Order Score Matching} \label{1st_sm}
For a given $\sigma$, we aim to learn a first-order score model $\mbf{s}_{\bsm{\theta}}(\cdot, \sigma): \mathbb{R}^N \rightarrow \mathbb{R}^N$ parameterized by $\bsm{\theta}$ with the following score matching objective:
\begin{align} 
	\min_{\bsm{\theta}} ~\mathbb{E}_{p(\tilde{\mbf{x}})}\left[ \left\|\mbf{s}_{\bsm{\theta}} (\tilde{\mbf{x}}, \sigma) - \nabla_{\tilde{\mbf{x}}} \log p(\tilde{\mbf{x}}) \right\|^2_2 \right].
\end{align}
The above formulation is shown equivalent to the denoising score matching objective \cite{vincent2011connection}, given by
\begin{align} \label{1st_dsm}
	\min_{\bsm{\theta}}~\mathbb{E}_{p(\mbf{x})p(\tilde{\mbf{x}}|\mbf{x})} \left[ \left\|\mbf{s}_{\bsm{\theta}} (\tilde{\mbf{x}}, \sigma) - \nabla_{\tilde{\mbf{x}}} \log p(\tilde{\mbf{x}}|\mbf{x}) \right\|^2_2 \right].
\end{align}
Substituting $\nabla_{\tilde{\mbf{x}}} \log p(\tilde{\mbf{x}}|\mbf{x}) = -\frac{\tilde{\mbf{x}} - \mbf{x}}{\sigma^2}$ into \eqref{1st_dsm} yields
\begin{align} \label{1st_dsm_2}
	\min_{\bsm{\theta}}~ \ell_1(\bsm{\theta}; \sigma) \triangleq \mathbb{E}_{p(\mbf{x})p(\tilde{\mbf{x}}|\mbf{x})} \left[ \left\|\mbf{s}_{\bsm{\theta}} (\tilde{\mbf{x}}, \sigma) + \frac{\tilde{\mbf{x}} - \mbf{x}}{\sigma^2} \right\|^2_2 \right].
\end{align}
In practice, the score-based MMSE denoiser is required to operate effectively across a range of noise levels.
We therefore combine \eqref{1st_dsm_2} for all possible $\sigma \in \{\sigma_i\}_{i=1}^L$ to formulate the following unified objective for training the first-order score model:
\begin{align}
	\min_{\bsm{\theta}} ~\mathcal{L}_1 \big(\bsm{\theta}; \{\sigma_i\}_{i=1}^L\big) \triangleq \frac{1}{L} \sum_{i=1}^{L} \lambda_1(\sigma_i)  \ell_1(\bsm{\theta}; \sigma_i),
\end{align}
where $\lambda_1(\sigma_i)$ is the weighting factor depending on $\sigma_i$.

\subsection{Second-Order Score Matching}
For a given $\sigma$, we aim to learn a second-order score model $\mbf{S}_{\bsm{\phi}}(\cdot, \sigma): \mathbb{R}^N \rightarrow \mathbb{R}^{N \times N}$ parameterized by $\bsm{\phi}$ with the following score matching objective:
\begin{align} \label{2nd_sm}
	\min_{\bsm{\phi}}~\mathbb{E}_{p(\tilde{\mbf{x}})} \left[ \left\|\mbf{S}_{\bsm{\phi}}(\tilde{\mbf{x}}, \sigma) - \nabla_{\tilde{\mbf{x}}}^2 \log p(\tilde{\mbf{x}}) \right\|^2_F \right].
\end{align}
The second-order score matching objective also has an equivalent denoising score matching formulation \cite{lu2022maximum}:
\begin{align} 
	\min_{\bsm{\phi}}~&\mathbb{E}_{p(\mbf{x})p(\tilde{\mbf{x}}|\mbf{x})} \bigg[ \Big\|\mbf{S}_{\bsm{\phi}}(\tilde{\mbf{x}}, \sigma) \nonumber \\
	&~~~~~~~~~~~~~~~- \mbf{b}(\mbf{x}, \tilde{\mbf{x}}, \sigma) \mbf{b}(\mbf{x},\tilde{\mbf{x}}, \sigma)^{\sf T} +  \frac{\mbf{I}}{\sigma^2}\Big\|_F^2\bigg], \label{2nd_dsm}
\end{align}
where $\mbf{b}(\mbf{x}, \tilde{\mbf{x}}, \sigma) \triangleq \nabla_{\tilde{\mbf{x}}} \log p(\tilde{\mbf{x}}) + (\tilde{\mbf{x}} - \mbf{x})/\sigma^2 $.
The objective in \eqref{2nd_dsm} relies on the ground truth of the first-order score function, which is in general not accessible. 
In practice, we replace $\nabla_{\tilde{\mbf{x}}} \log p(\tilde{\mbf{x}})$ by the learned $\mbf{s}_{\bsm{\theta}}(\tilde{\mbf{x}}, \sigma)$ for efficient computation.


As indicated in \eqref{post_variance}, only the trace of the posterior covariance matrix is needed, rather than the covariance matrix itself.
By noting in \eqref{2nd_order_tweedie} the simple relation between the posterior covariance matrix and the second-order score function, 
it suffices to only match the trace of the second-order score function for posterior variance evaluation.
The simplified objective is expressed as
\begin{align} \label{2nd_dsm_trace}
	\min_{\bsm{\phi}}~ \ell_2(\bsm{\phi}; \sigma) &\triangleq \mathbb{E}_{p(\mbf{x})p(\tilde{\mbf{x}}|\mbf{x})} \bigg[ \bigg|\mathrm{tr}\big(\mbf{S}_{\bsm{\phi}}(\tilde{\mbf{x}}, \sigma)\big) \nonumber \\
	&~~~~~~~~~~~~~~~- \left\|\hat{\mbf{b}}(\mbf{x}, \tilde{\mbf{x}}, \sigma)\right\|_2^2 +  \frac{N}{\sigma^2}\bigg|^2\bigg],
\end{align}
where $\hat{\mbf{b}}(\mbf{x}, \tilde{\mbf{x}}, \sigma) \triangleq \mbf{s}_{\bsm{\theta}}(\tilde{\mbf{x}}, \sigma) + (\tilde{\mbf{x}} - \mbf{x})/\sigma^2$.
We combine \eqref{2nd_dsm_trace} for all $\sigma \in \{\sigma_i\}_{i=1}^L$ to obtain the following unified objective for training the second-order score model:
\begin{align} \label{2nd_dsm_unified}
	\min_{\bsm{\phi}}~\mathcal{L}_2 \big(\bsm{\phi}; \{\sigma_i\}_{i=1}^L\big) \triangleq \frac{1}{L} \sum_{i=1}^{L} \lambda_2(\sigma_i)  \ell_2(\bsm{\phi}; \sigma_i),
\end{align}
where $\lambda_2(\sigma_i)$ is the weighting factor depending on $\sigma_i$.

In practice, one can first train a first-order score model, and then freeze it when applied to second-order score learning.
Alternatively, the two score models can be trained both at once by disabling gradient backpropagation from the second-order training objective \eqref{2nd_dsm_unified} to the first-order score model.

\section{STMP Algorithm}
The score models, once learned, can be plugged into the Turbo-CS framework, facilitating MMSE denoising via
\begin{align}
	\mbf{x}_{\sf B}^\textit{post} &= \mbf{x}_{\sf B}^\textit{pri} + v_{\sf B}^\textit{pri} \mbf{s}_{\bsm{\theta}} \big(\mbf{x}_{\sf B}^\textit{pri}, \sqrt{v_{\sf B}^\textit{pri}}\big), \label{posterior_mean_stmp} \\
	v_{\sf B}^\textit{post} &= \frac{(v_{\sf B}^\textit{pri})^2}{N} \mathrm{tr} \Big(\mbf{S}_{\bsm{\phi}} \big(\mbf{x}_{\sf B}^\textit{pri}, \sqrt{v_{\sf B}^\textit{pri}}\big) \Big) + v_{\sf B}^\textit{pri}. \label{posterior_variance_stmp}
\end{align}
Eqns. \eqref{posterior_mean_stmp} and \eqref{posterior_variance_stmp} are simply restatements of \eqref{1st_order_tweedie} and \eqref{2nd_order_tweedie}, respectively. 
Upon this replacement, we obtain the STMP algorithm in Algorithm \ref{alg:score_turbo_mp}.
The algorithm stops when the maximum number of iterations is reached, or the difference of $\mbf{x}_{\sf B}^\textit{post}$ between two consecutive iterations falls below a certain threshold.

We have empirically observed that when the compression ratio $M/N$ is low, adding damping to the message updates helps to stabilize convergence. 
In particular, we suggest to replace Line \ref{to_be_damped_A} of Algorithm \ref{alg:score_turbo_mp} by
\begin{align}
	\mbf{x}_{\sf B}^\textit{pri} &= \beta \mbf{x}_{\sf A}^\textit{ext} + (1-\beta) \mbf{x}_{{\sf A}, \textit{old}}^\textit{ext}, \\
	v_{\sf B}^\textit{pri} &= \beta v_{\sf A}^\textit{ext} + (1-\beta) v_{{\sf A},\textit{old}}^\textit{ext},
\end{align}
where $\mbf{x}_{{\sf A}, \textit{old}}^\textit{ext}$ and $v_{{\sf A},\textit{old}}^\textit{ext}$ are the extrinsic mean and variance of module A from the last iteration, $\beta \in (0, 1]$ is the damping factor.
Line \ref{to_be_damped_B} of Algorithm \ref{alg:score_turbo_mp} should be modified similarly.

As shown in Lines \ref{stmp_module_a_mean} and \ref{stmp_module_b_variance} of Algorithm \ref{alg:score_turbo_mp}, the LMMSE estimator requires the computation of a different matrix inverse at each iteration with prohibitive complexity.
This can be avoided by the pre-calculation of the singular value decomposition (SVD) of $\mbf{A}$ \cite{vamp}.
Moreover, in compressive imaging, the rows of $\mbf{A}$ are typically randomly selected from an orthogonal matrix (e.g., discrete cosine transform (DCT) matrix).
This results in a diagonal matrix to be inverted at each iteration, which significantly reduces the computational cost.

\begin{algorithm}[t]
	\caption{STMP Algorithm}
	\label{alg:score_turbo_mp}
	\begin{algorithmic}[1]
		\STATE {\bfseries Input:} $\mbf{A}$, $\mbf{y}$, $\delta_0^2$, $\mbf{x}_{\sf A}^\textit{pri}$, $v_{\sf A}^\textit{pri}$
		\STATE {\bfseries Output:} $\mbf{x}_{\sf B}^\textit{post}$
		\REPEAT
		\STATE \% LMMSE estimator
		\STATE $\mbf{x}_{\sf A}^{\textit{post}} \!= \! \mbf{x}_{\sf A}^\textit{pri} \!+\! v_{\sf A}^\textit{pri}\mbf{A}^{\sf T} \big(v_{\sf A}^\textit{pri}\mbf{A}\mbf{A}^{\sf T} \!+\! \delta_0^2\mbf{I}\big)^{-1} \!\big(\mbf{y} \! - \! \mbf{A} \mbf{x}_{\sf A}^\textit{pri}\big)$ \label{stmp_module_a_mean}
		\STATE $v_{\sf A}^\textit{post} = v_{\sf A}^\textit{pri} - \frac{(v_{\sf A}^\textit{pri})^2}{N} \mathrm{tr} \big(\mbf{A}^{\sf T} \big(v_{\sf A}^\textit{pri}\mbf{A} \mbf{A}^{\sf T} + \delta_0^2 \mbf{I}\big)^{-1}\mbf{A}\big)$ \label{stmp_module_b_variance}
		\STATE $v_{\sf A}^{\textit{ext}} = 1/\big(1/v_{\sf A}^{\textit{post}} - 1/v_{\sf A}^{\textit{pri}}\big)$ 
		\STATE $\mbf{x}_{\sf A}^{\textit{ext}} = v_{\sf A}^{\textit{ext}} \big(\mbf{x}_{\sf A}^{\textit{post}}/v_{\sf A}^{\textit{post}} - \mbf{x}_{\sf A}^{\textit{pri}}/v_{\sf A}^{\textit{pri}}\big)$ 
		\STATE $\mbf{x}_{\sf B}^\textit{pri} = \mbf{x}_{\sf A}^\textit{ext}, v_{\sf B}^\textit{pri} = v_{\sf A}^\textit{ext}$ \label{to_be_damped_A}
		\STATE \% Score-based MMSE denoiser
		\STATE $\mbf{x}_{\sf B}^\textit{post} = \mbf{x}_{\sf B}^\textit{pri} + v_{\sf B}^\textit{pri} \mbf{s}_{\bsm{\theta}} \big(\mbf{x}_{\sf B}^\textit{pri}, \sqrt{v_{\sf B}^\textit{pri}}\big)$ 
		\STATE $v_{\sf B}^\textit{post} = \frac{(v_{\sf B}^\textit{pri})^2}{N} \mathrm{tr} \Big(\mbf{S}_{\bsm{\phi}} \big(\mbf{x}_{\sf B}^\textit{pri}, \sqrt{v_{\sf B}^\textit{pri}}\big) \Big) + v_{\sf B}^\textit{pri}$ 
		\STATE $ v_{\sf B}^{\textit{ext}} = 1/\big(1/v_{\sf B}^{\textit{post}} - 1/v_{\sf B}^{\textit{pri}}\big)$ 
		\STATE $ \mbf{x}_{\sf B}^{\textit{ext}} = v_{\sf B}^{\textit{ext}} \big(\mbf{x}_{\sf B}^{\textit{post}}/v_{\sf B}^{\textit{post}} - \mbf{x}_{\sf B}^{\textit{pri}}/v_{\sf B}^{\textit{pri}}\big)$ 
		\STATE $\mbf{x}_{\sf A}^\textit{pri} = \mbf{x}_{\sf B}^\textit{ext}, v_{\sf A}^\textit{pri} = v_{\sf B}^\textit{ext}$ \label{to_be_damped_B}
		\UNTIL{the stopping criterion is met}
	\end{algorithmic}
\end{algorithm}

\section{Experiments}
In this section, we evaluate the performance of the proposed method based on the FFHQ $256\times 256$ dataset \cite{ffhq}.
We adopt the pre-trained score function in \cite{score_sde} as our first-order score model.
Unfortunately, there is no publicly available second-order score model on the FFHQ dataset. 
For training the second-order score model, we adopt the NCSN++ architecture \cite{score_sde} that outputs only the diagonal value of the second-order score, yielding the same output dimension as the first-order score model.
The training of the second-order score model follows the same default settings as its first-order counterpart \cite{score_sde}. 
For inference, we test our proposed method and several benchmarks on $1000$ images from the validation set of the FFHQ dataset.

We consider a compressive image recovery task with the measurement matrix in the form of $\mbf{A} = \mbf{S} \mbf{W} \bsm{\Theta}$, where $\mbf{S} \in \mathbb{R}^{M \times N}$ is a random row selection matrix consisting of randomly selected rows from a permutation matrix, $\mbf{W} \in \mathbb{R}^{N \times N}$ is a DCT matrix, and $\bsm{\Theta} \in \mathbb{R}^{N \times N}$ is a diagonal matrix with random signs ($1$ or $-1$) in the diagonal.
We provide comprehensive results on the image recovery performance under different compression ratios and noise levels. 

We compare our proposed method against both the recent state-of-the-art score-based posterior sampling techniques and the traditional plug-and-play methods.
The baselines include diffusion posterior sampling (DPS) \cite{dps}, diffusion models for plug-and-play image restoration (DiffPIR) \cite{diffpir}, plug-and-play alternating direction method of multipliers (PnP-ADMM) \cite{pnp_admm}, and D-Turbo-CS \cite{xue2017access}.
As suggested in \cite{dps, diffpir}, 
we set the number reverse timesteps (i.e., NFEs) in DPS and DiffPIR as $1000$ and $100$, respectively.
We choose the total variation (TV) denoising method as the plug-and-play denoiser in PnP-ADMM, and choose the SURE-LET \cite{sure-let} denoiser for D-Turbo-CS.

\begin{figure}[t]
	\centering
	\vspace{-.5em}
	\begin{minipage}{0.493\linewidth}
		\centering		
		\includegraphics[width=\linewidth]{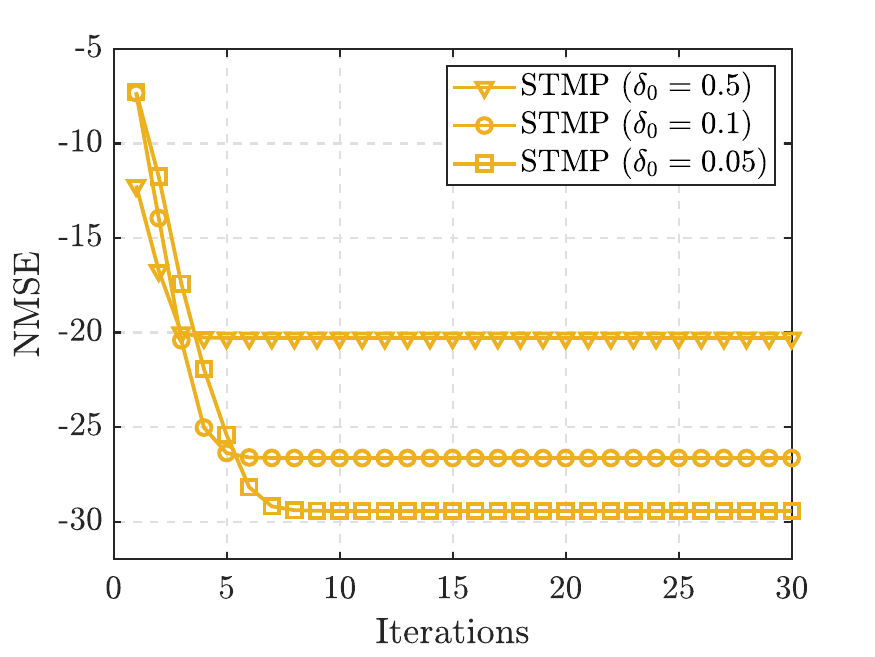}\\
		\scriptsize{(a) $M/N = 0.5$, $\beta = 0.8$}
	\end{minipage}
	\begin{minipage}{0.493\linewidth}
		\centering		
		\includegraphics[width=\linewidth]{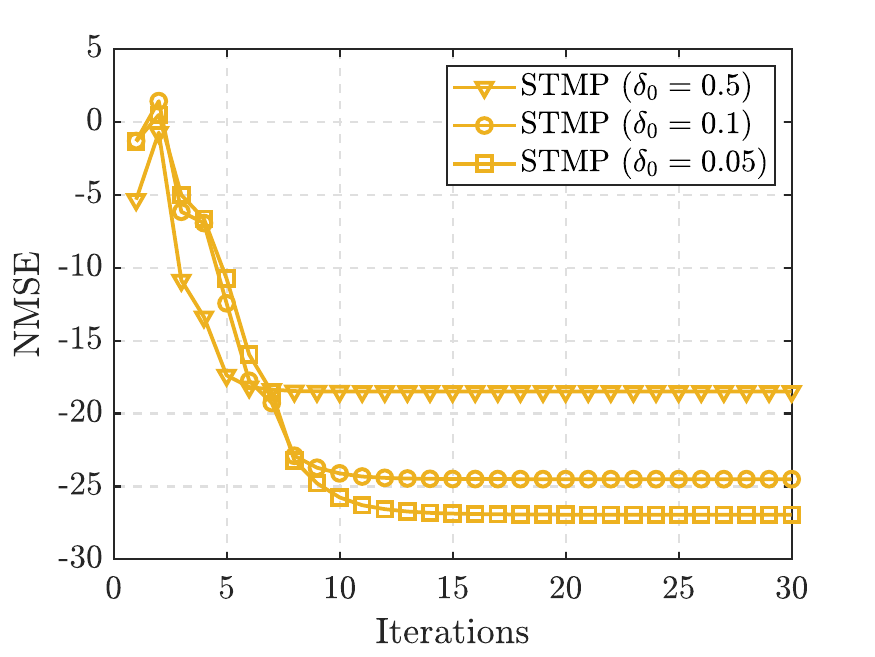}\\
		\scriptsize{(b) $M/N = 0.2$, $\beta = 0.6$}
	\end{minipage}
	\caption{\label{fig_convergence}
		Convergence behaviors of STMP for different compression ratios and noise levels.}
\end{figure}

Fig. \ref{fig_convergence} reports the normalized MSE (NMSE) of the reconstruction during the execution of STMP.
Overall, the proposed algorithm exhibits a fast convergence speed under different compression ratios $M/N$ and noise levels $\delta_0$.
The number of iterations required to reach convergence increases as the ratio $M/N$ decreases, since recovering from fewer observations is more ill-posed and challenging.
Moreover, the proposed algorithm converges slightly slower with a small added noise.
It recovers more details of the ground-truth image from relatively ``clean'' observations. 

\newcommand{\imwidth}{1.37cm}
\begin{figure}
	\centering
	\def\arraystretch{0.7}
	\setlength\tabcolsep{0.03cm}
	\begin{tabular}{lcccccc}
		\multirow{2}{*}{\raisebox{-0.7cm}[0pt][0pt]{\rotatebox{90}{\scriptsize $\delta_0 = 0.05$}}} &
		\includegraphics[width=\imwidth,height=\imwidth]{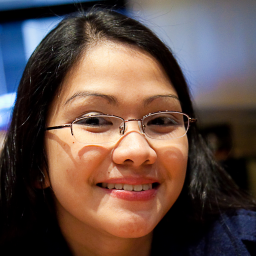}
		& \includegraphics[width=\imwidth,height=\imwidth]{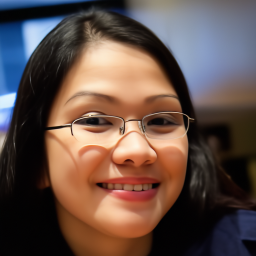}
		& \includegraphics[width=\imwidth,height=\imwidth]{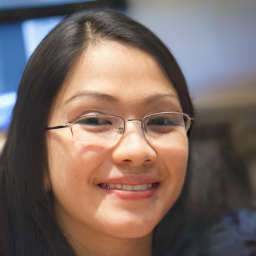}
		& \includegraphics[width=\imwidth,height=\imwidth]{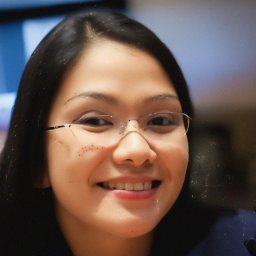}
		& \includegraphics[width=\imwidth,height=\imwidth]{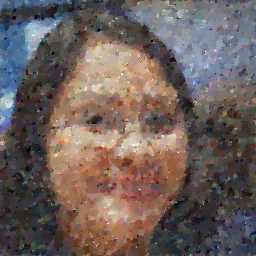}
		& \includegraphics[width=\imwidth,height=\imwidth]{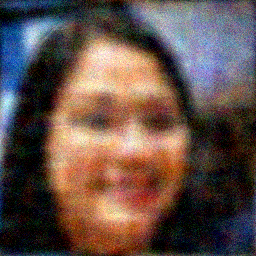}  \\
		&
		\includegraphics[width=\imwidth,height=\imwidth]{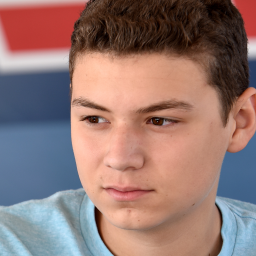}
		& \includegraphics[width=\imwidth,height=\imwidth]{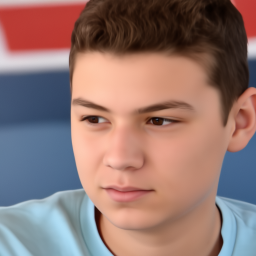}
		& \includegraphics[width=\imwidth,height=\imwidth]{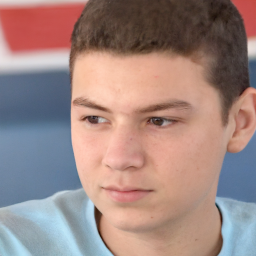}
		& \includegraphics[width=\imwidth,height=\imwidth]{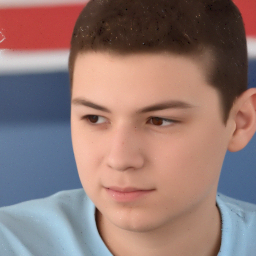}
		& \includegraphics[width=\imwidth,height=\imwidth]{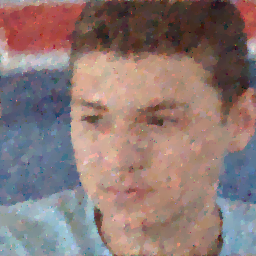}
		& \includegraphics[width=\imwidth,height=\imwidth]{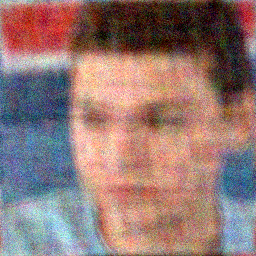} \\
		\multicolumn{7}{c}{\vspace{-.2em}} \\
		\multirow{2}{*}{\raisebox{-0.55cm}[0pt][0pt]{\rotatebox{90}{\scriptsize $\delta_0 = 0.5$}}} &
		\includegraphics[width=\imwidth,height=\imwidth]{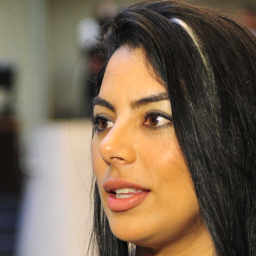}
		& \includegraphics[width=\imwidth,height=\imwidth]{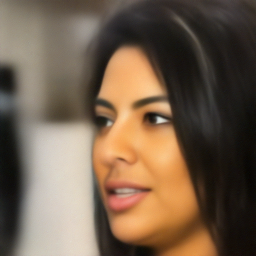}
		& \includegraphics[width=\imwidth,height=\imwidth]{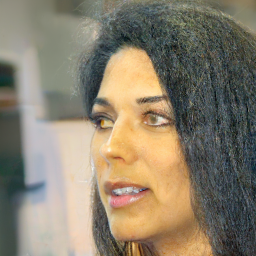}
		& \includegraphics[width=\imwidth,height=\imwidth]{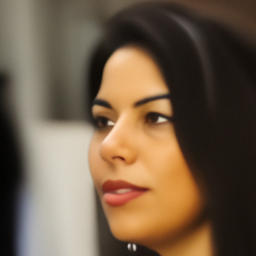}
		& \includegraphics[width=\imwidth,height=\imwidth]{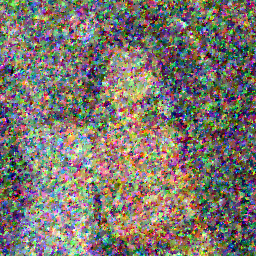}
		& \includegraphics[width=\imwidth,height=\imwidth]{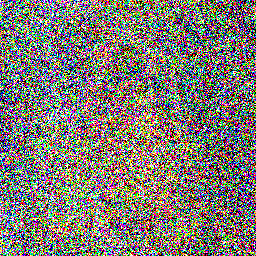} \\
		&
		\includegraphics[width=\imwidth,height=\imwidth]{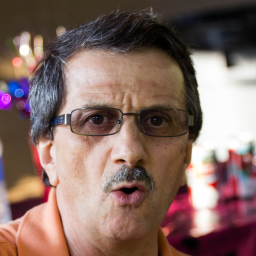}
		& \includegraphics[width=\imwidth,height=\imwidth]{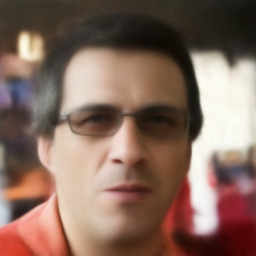}
		& \includegraphics[width=\imwidth,height=\imwidth]{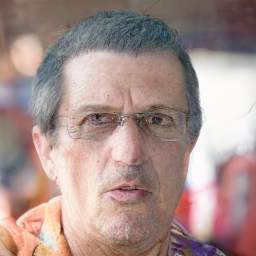}
		& \includegraphics[width=\imwidth,height=\imwidth]{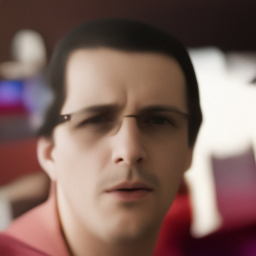}
		& \includegraphics[width=\imwidth,height=\imwidth]{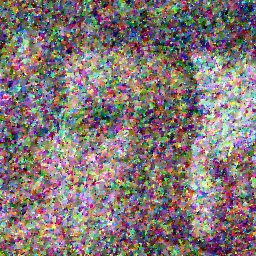}
		& \includegraphics[width=\imwidth,height=\imwidth]{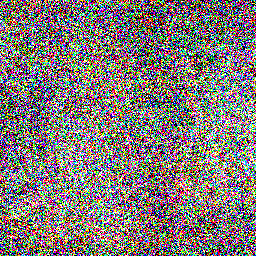}  \\
		& \scriptsize Ground truth & \scriptsize STMP & \scriptsize DPS & \scriptsize DiffPIR & \scriptsize PnP-ADMM & \scriptsize D-Turbo-CS
	\end{tabular}
	\caption{Representative results on FFHQ validation set. The compression ratio is $M/N = 0.1$ and we set $\beta = 0.5$.}
	\label{fig:imagenet_color}
	\vspace{-.8em}
\end{figure}

We present some representative results in Fig. \ref{fig:imagenet_color}.
Both PnP-ADMM and D-Turbo-CS fail to produce satisfactory results due to the extremely low compression ratio ($M/N=0.1$).
When the added noise is small ($\delta_0 = 0.05$), the proposed method, DPS, and DiffPIR are able to generate high-quality reconstructions.
The DPS baseline, however, does not always adhere to the ground-truth image when the added noise is large ($\delta_0 = 0.5$).
For example, in the last row of Fig. \ref{fig:imagenet_color}, the sample generated by DPS depicts a man wearing transparent glasses, whereas the ground truth shows sunglasses.
This highlights the superior restoration fidelity of our method over DPS. 
In Table \ref{tab:in1k_noisy},
we report the average peak signal-to-noise ratio (PSNR), structural similarity index measure (SSIM), Fréchet inception distance (FID), and learned perceptual image patch similarity (LPIPS) to evaluate the faithfulness to the original image and the visual quality.
We see that our proposed algorithm outperforms other benchmarks on most of these metrics.

\begin{table*}[t]
	\centering
	\caption{Quantitative results on FFHQ validation set. The compression ratio is $M/N = 0.1$ and we set $\beta=0.5$.}
	\vspace{-1.2em}
	\label{tab:in1k_noisy}
	\begin{center}
		\begin{tabular}{l p{0.0001\textwidth} cccc p{0.0001\textwidth} cccc}
			\toprule
			\multirow{2}{*}{Method} & ~ &  \multicolumn{4}{c}{$\delta_0 = 0.5$} & ~ & \multicolumn{4}{c}{$\delta_0 = 0.05$} \\
			\cline{3-6} \cline{8-11} 
			& ~ & PSNR$\uparrow$ & SSIM$\uparrow$ & FID$\downarrow$ & LPIPS$\downarrow$ & ~ & PSNR$\uparrow$ & SSIM$\uparrow$ & FID$\downarrow$ & LPIPS$\downarrow$ \\
			\midrule
			STMP (Ours) & ~ & $\mathbf{22.07}$ & $\mathbf{0.6447}$ & $\underline{68.85}$ & $\mathbf{0.0640}$ & ~ & $\mathbf{29.86}$ & $\mathbf{0.8615}$ & $\underline{46.09}$ & $\mathbf{0.0146}$ \\
			DPS  & ~ & $14.01$ & $0.4447$ & $\mathbf{56.23}$ & $0.1430$ & ~ & $22.15$ & $0.7170$ & $\mathbf{29.36}$ & $0.0287$ \\
			DiffPIR & ~ & $\underline{21.07}$ & $\underline{0.6149}$ & $97.38$ & $\underline{0.0765}$ & ~ & $\underline{25.52}$ & $\underline{0.7539}$ & $56.30$ & $\underline{0.0270}$ \\
			PnP-ADMM  & ~ & $11.46$ & $0.0638$ & $374.52$ & $0.4526$ & ~ & $19.16$ & $0.4646$ & $310.09$ & $0.0827$ \\
			D-Turbo-CS & ~ & $5.77$ & $0.0045$ & $427.89$ & $0.9031$ & ~ & $19.15$ & $0.2072$ & $251.94$ & $0.1558$ \\  
			\bottomrule
		\end{tabular}
	\end{center}
	\vspace{-.5em}
\end{table*}

\begin{figure}[t]
	\centering
	\vspace{-.1in}
	\begin{minipage}{0.493\linewidth}
		\centering
		\includegraphics[width=\linewidth]{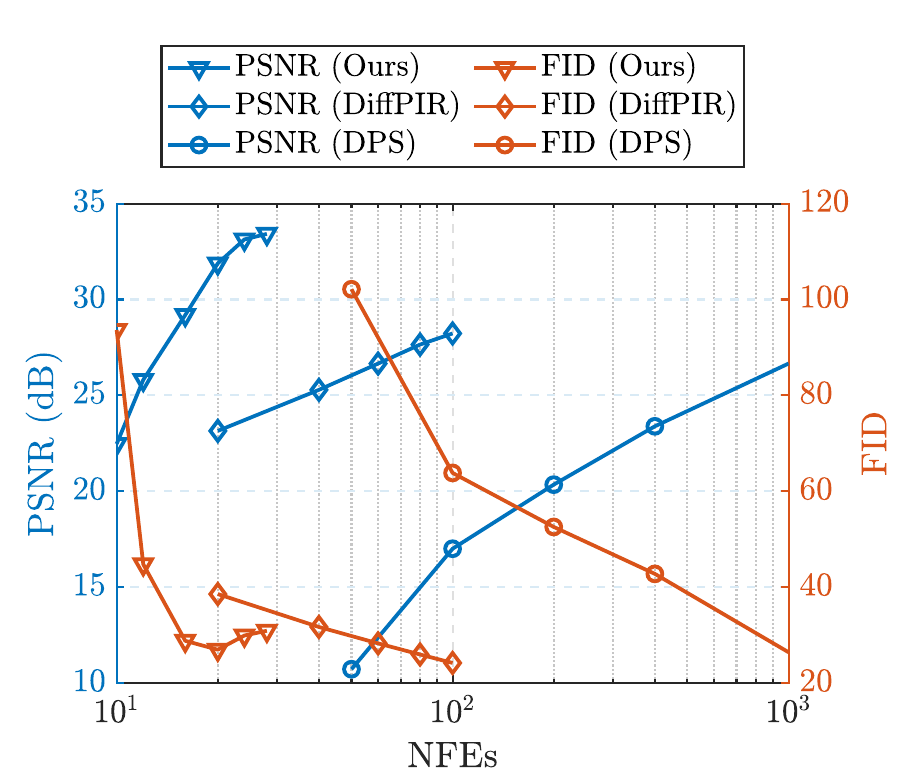}\\
		\scriptsize{(a) PSNR/FID vs. NFEs}
	\end{minipage}
	\begin{minipage}{0.493\linewidth}
		\centering		
		\includegraphics[width=\linewidth]{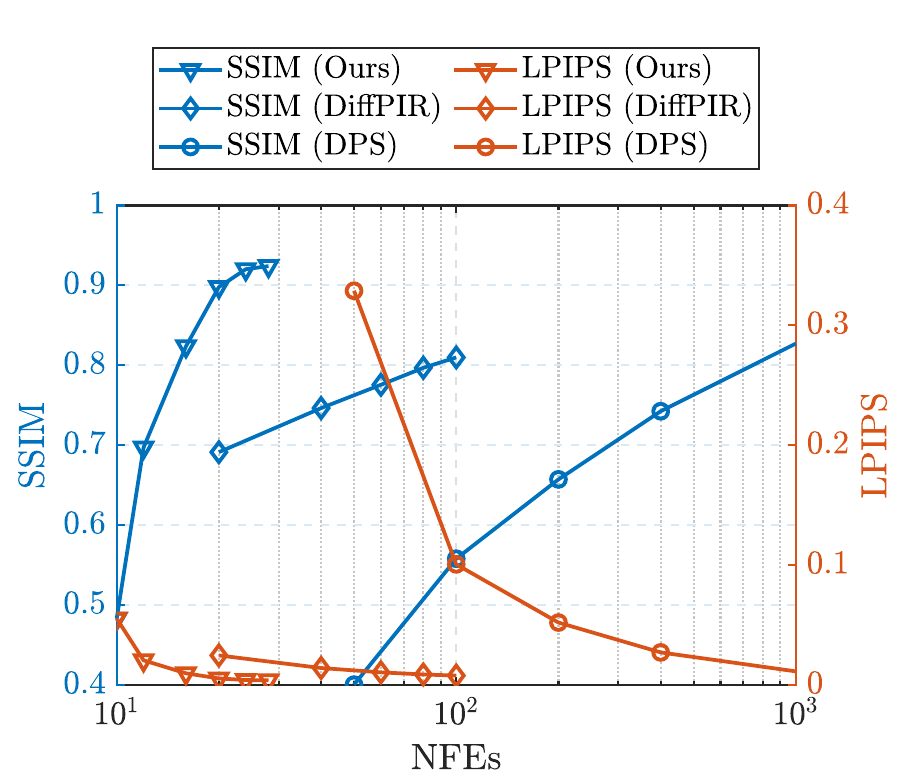}\\
		\scriptsize{(b) SSIM/LPIPS vs. NFEs}
	\end{minipage}
	\caption{\label{fig_tradeoff}
		Tradeoff between the reconstruction faithfulness/image quality and the computational complexity. The compression ratio is $M/N=0.4$ and the noise level is $\delta_0 = 0.05$. We set $\beta = 0.8$.}
	\vspace{-.09in}
\end{figure}

The computational complexity of STMP, DPS, and DiffPIR is dominated by the number of NFEs.
In STMP, each iteration involves forward propagation through both the first- and the second-order score networks.
Although these two NFEs can be executed in parallel without doubling the time consumption, we still consider each iteration as two NFEs.
We evaluate the tradeoff between the reconstruction faithfulness/image quality and the computational complexity in Fig. \ref{fig_tradeoff}.
Compared to the other two benchmarks, STMP requires significantly fewer NFEs to produce meaningful results, highlighting the efficiency of our design.
Note that DPS requires not only forward propagation during each NFE, but also back-propagating gradients through the score network.
This makes the sampling steps in DPS even more computationally expensive.


\section{Conclusion}
In this paper, we introduced score-based turbo message passing (STMP), a novel approach for compressive image recovery relying on score-based generative modeling.
STMP inherits the fast convergence speed of message passing algorithms and, meanwhile, takes full advantage of the expressive prior through the integration of score networks. 
Experimental results on the FFHQ dataset showcase the state-of-the-art performance of STMP against different benchmarks.

\bibliographystyle{IEEEtran}
\bibliography{IEEEabrv,spawc_bib}

\end{document}